\documentclass[authoryear,final,1p,times,twocolumn]{elsarticle}

\usepackage{amssymb}
\usepackage{amsmath}

\usepackage{xcolor}
\usepackage{multirow}
\usepackage{tabularx}
\usepackage{booktabs}
\usepackage{array}
\usepackage{makecell}
\usepackage{pifont}
\usepackage[capitalise, noabbrev]{cleveref}
\newcommand*\revise[1]{{#1}}

\journal{Journal}

\begin{document}

\begin{frontmatter}


\cortext[cor1]{Corresponding author: Xiaojuan Ma}

\title{The Speculative Future of Conversational AI for Neurocognitive Disorder Screening: a Multi-Stakeholder Perspective}


\author[label1]{Jiaxiong Hu}
\ead{hujx@ust.hk}

\author[label1-1,label2]{Ruowen Niu}
\ead{rniu440@connect.hkust-gz.edu.cn}

\author[label5,label1]{Qiuxin Du}
\ead{qiuxin.du.25@ucl.ac.uk}

\author[label3,label2]{Chenzhuo Xiang}
\ead{xcz24@mails.tsinghua.edu.cn}

\author[label4,label2]{Yirui Zuo}
\ead{24s034097@stu.hit.edu.cn}

\author[label2]{Jihong Jeung}
\ead{jihong95@tsinghua.edu.cn}

\author[label1]{Xiaojuan Ma\corref{cor1}}
\ead{mxj@cse.ust.hk}

\affiliation[label1]{organization={Department of Computer Science and Engineering, the Hong Kong University of Science and Technology},
            city={Hong Kong SAR},
            country={China}}

\affiliation[label1-1]{organization={Department of Computational Media and Arts, the Hong Kong University of Science and Technology (Guangzhou)},
            city={Guangzhou},
            postcode={511453},
            state={Guangdong},
            country={China}}
            
\affiliation[label2]{organization={The Future Laboratory, Tsinghua University},
            city={Beijing},
            postcode={100084}, 
            country={China}}

\affiliation[label3]{organization={Academy of Arts \& Design, Tsinghua University},
            city={Beijing},
            postcode={100084}, 
            country={China}}

\affiliation[label4]{organization={Harbin Institute of Technology},
            city={Harbin},
            postcode={150001}, 
            state={Heilongjiang},
            country={China}}
            
\affiliation[label5]{organization={UCL Interaction Center, University College London},
            city={London},
            country={the United Kingdom}}

\begin{abstract}
Neurocognitive disorders (NCDs), such as Alzheimer's disease, are globally prevalent and require scalable screening methods for proactive management. Prior research has explored the potential of technologies like conversational AI (CAI) to administer NCD screening tests. However, challenges remain in designing CAI-based solutions that make routine NCD screening socially acceptable, engaging, and capable of encouraging early medical consultation. In this study, we conducted interviews with 36 participants, including clinicians, individuals at risk of NCDs, and their caregivers, to explore the speculative future of adopting CAI for NCD screening. Our findings reveal shared expectations, such as deploying CAI in home or community settings to reduce social stress. Nonetheless, conflicts emerged among stakeholders, for example, users' need for emotional support may conflict with clinicians' preference for CAI's professional and standardized administration. Then, we look into the user journey of NCD screening based on the current practice of manual screening and the expected CAI-supported screening. Finally, leveraging the human-centered approach, we provide actionable implications for future CAI design in NCD screening.
\end{abstract}







\end{frontmatter}

\section{Introduction}
Neurocognitive disorders (NCDs), including those associated with Alzheimer's disease, exert a profound negative impact on the quality of life for both patients and their caregivers~\citep{patterson2022world}. 
The rising prevalence of NCDs constitutes a critical global public health concern~\citep{Association2024AD, rajan2021population}.
While NCDs are typically regarded as incurable and irreversible~\citep{zverova2019Clinical}, emerging therapeutics such as Lecanemab have demonstrated efficacy when administered during the early stages of the disease~\citep{chris2023lecanemab}. 
This underscores the pivotal role of routine NCD screening in facilitating early interventions to mitigate disease progression~\citep{Akpan2019Neurocognitive, ferri2005global, du2024lightsword}, thereby ultimately enhancing the quality of life for affected individuals and their caregivers.
However, numerous countries are grappling with similar barriers in scaling up NCD screening, stemming from rapidly aging populations, shortages of specialized clinical personnel, and limited access to early-stage assessment~\citep{long2018barriers}.
China represents a particularly salient context for examining these challenges.
As one of the world's most rapidly aging societies, it faces a profound demographic shift toward population aging, alongside substantial barriers to neurocognitive disorder screening.
Screening services remain largely hospital-centered, resource-intensive, and geographically unevenly distributed, which limits early detection and timely, targeted interventions~\citep{jia2020prevalence, nichols2022estimation}.
Notably, these barriers are not exclusive to China. 
Evidence from other Asian regions~\citep{jitapunkul2009undiagnosed} and the Americas~\citep{liu2024Detection} indicates the presence of analogous structural obstacles in diverse contexts. 
Accordingly, we position China as a representative case to explore actionable solutions addressing NCD screening barriers in aging populations globally.
Conversational artificial intelligence (CAI) exhibits significant potential for the interactive administration of complex NCD screening tasks on behalf of clinicians.
Prior Human-Computer Interaction (HCI) research has explored CAI-administered NCD screening employing human-centered methodologies, including user experience evaluation~\citep{ding2022talktive, hu2024designing}.
Cutting-edge AI techniques, most notably large language models (LLMs), have been leveraged to augment CAI's conversational capabilities.
However, administering tests constitutes merely one component of comprehensive NCD screening and management.
A critical challenge persists in designing CAI solutions that render routine NCD screening socially acceptable, engaging, and capable of facilitating users' decision-making processes regarding timely medical consultation.
Prior studies underscore that attaining this objective necessitates multi-stakeholder collaboration~\citep{callahan2006effectiveness, heintz2020Emerging, frost2021Implementing}.
Accordingly, an in-depth exploration of user needs and expectations from a multi-stakeholder perspective is imperative to inform the future design of CAI-supported NCD screening systems.

Researchers have proposed various models and frameworks to delineate the landscape of collaborative care for NCD patients~\citep{heintz2020Emerging}; however, the intricate multi-stakeholder dynamics inherent in NCD screening remain underexplored.
Furthermore, the integration of emerging technologies such as CAI into existing workflows inevitably necessitates adaptations to both collaborative processes and the mindsets of participating stakeholders~\citep{dolgikh2021collaborative}.
This transitional process requires a comprehensive understanding of how NCD screening can be effectively implemented and coordinated among key stakeholder groups, as well as the socioeconomic, technical, and organizational barriers associated with scaling up CAI adoption for NCD screening purposes.
To address these research gaps, the present study conducts a qualitative investigation in China.
Building upon existing research on multi-stakeholder collaborative interventions for patients with dementia~\citep{sun2024TechnologyMediated} and rooted in the practical context of NCD screening practices in China~\citep{wang2019continuum, luo2025evolution}, this study focuses on three key stakeholder groups: clinicians, individuals at risk of NCDs, and their caregivers.
Clinicians are defined as healthcare professionals practicing in hospital departments pertinent to NCD care (e.g., neurology or geriatrics) who are tasked with conducting professional diagnosis~\citep{van2024computerized}.
Individuals at risk of NCDs refer to those aged 60 years and above, given that the prevalence of NCDs increases substantially from this age onward in China~\citep{jia2020prevalance}.
Finally, caregivers in this study are specifically defined as unpaid family members, as family-centered caregiving remains the dominant model for NCD care in China~\citep{li2022factors}.
Caregivers play an important role in identifying early signs of NCDs, proactively initiating screening, and supporting older adults throughout the screening process, particularly when stigma and inadequate understanding of NCDs lead to hesitancy in participation~\citep{wang2019continuum}.

In the present study, semi-structured interviews were conducted with the three aforementioned key stakeholder groups.
Guided by the design methodology of ``Alternative Presents and Speculative Futures''~\citep{auger2010alternative}, this study explored and speculated about the near future of scalable CAI-enabled NCD screening.
A total of 36 participants were recruited from seven cities across China, comprising 5 clinicians, 7 caregivers, and 24 individuals at risk of NCDs (11 of whom reported subjective cognitive decline or had previously been diagnosed with mild NCDs).
To facilitate participants' comprehension and assist them in envisioning future scenarios, video clips depicting NCD screening procedures administered by both human professionals and CAI were presented.
The interview findings uncovered both convergent and divergent expectations among the stakeholder groups.
For instance, individuals at risk of NCDs and caregivers expressed a preference for CAI to provide immediate diagnostic outputs, whereas clinicians stressed that CAI should be limited to offering recommendations for seeking medical consultation instead of a formal diagnosis.
Additionally, the current workflow of NCD screening was delineated, and barriers inherent in existing screening practices, such as social distress, were elucidated.
This work advances a human-centered understanding of CAI adoption for NCD screening in a speculative future, from the perspective of multiple key stakeholder groups.
First, through semi-structured interviews with multi-stakeholder groups, we identified convergences and divergences in their expectations toward CAI-supported NCD screening.
Second, we derived user journey maps for NCD screening from interview accounts to provide a holistic understanding of multi-stakeholder user pain points and the potential benefits and constraints of CAI integration.
Third, leveraging a human-centered research approach, we put forward actionable design implications for future CAI-supported NCD screening, aimed at facilitating the timely detection and intervention of NCDs.

\section{Related Works}

\subsection{Neurocognitive Disorder Screening}
In recent years, global population aging has resulted in a continuous increase in the number of patients diagnosed with NCDs. 
By 2050, the number of NCD patients worldwide was projected to triple, reaching 152 million~\citep{patterson2018world}.
As one of the countries most affected by Alzheimer's disease and related NCDs~\citep{nichols2022estimation}, the number of NCD patients aged 65 and above in the United States was projected to increase from 6.9 million in 2021 to 13.8 million by 2060~\citep{rajan2021population}.
According to a report in 2022~\citep{ren2022china}, China accounted for approximately one-quarter of the global population affected by NCDs, with 15.07 million NCD patients aged 60 and above and 38.77 million individuals diagnosed with mild NCDs~\citep{jia2020prevalence}. 
By 2050, the number of NCD cases in China was projected to rise to 45.53 million~\citep{nichols2022estimation}.

While NCD-related diseases such as Alzheimer's disease are generally declared incurable and irreversible~\citep{zverova2019Clinical}, new drugs such as Lecanemab were tested as effective for reducing cognitive and functional decline in individuals with early NCD only~\citep{chris2023lecanemab}. 
Therefore, screening for NCDs is crucial for implementing early interventions to help slow the progression of the disease~\citep{Akpan2019Neurocognitive, ferri2005global, du2024lightsword}, improving the quality of life of affected individuals and their caregivers.
To achieve early screening and diagnosis, physicians and clinicians use tests like the Mini-Mental State Examination (MMSE)~\citep{flstein1983MMSE}, Montreal Cognitive Assessment (MoCA)~\citep{Nasreddine2005Montreal}, and Hong Kong Grocery Shopping Dialog Task (HK-GSDT)~\citep{gong2022hongkong}, to evaluate various cognitive functions, such as memory, attention, and language skills. 
However, these methods require in-person administration, limiting their scalability to large populations~\citep{long2018barriers}.


NCDs are frequently not identified in a timely manner~\citep{Association2024AD}. 
A recent study highlighted this issue, revealing that merely 8\% of older Americans with mild NCDs received a formal diagnosis~\citep{liu2024Detection}. 
Globally, it was estimated that 75\% of individuals with NCDs remained undiagnosed. 
In low- and middle-income countries, this figure rises to 90\% due to the stigma surrounding NCDs and a lack of awareness~\citep{gauthier2021world}.  
Similar under-detection and delayed diagnosis have also been reported among Asian populations, where cultural beliefs, stigma, and limited awareness influence the recognition and disclosure of cognitive impairment~\citep{Jang2010Asian}.
In China, 93\% of dementia were undetected~\citep{Chen2013determinants}, while in Thailand, the rate was 95.6\%~\citep{jitapunkul2009undiagnosed}. 
Overall, the existing diagnostic methods are still insufficient for the early detection of a vast number of patients.

\subsection{Multi-stakeholder Collaboration for NCD Care}

As NCD prevalence increases, there is a growing need for multi-stakeholder collaboration to support individuals with NCDs and their families~\citep{Association2024AD}.
Primary care physicians, although the first point of contact for patients, frequently lack the necessary time, resources, and training to effectively screen for, diagnose, and manage NCDs~\citep{heintz2020Emerging}.
Extensive research over the past several decades highlights the effectiveness of collaborative primary care models in treating individuals with NCDs~\citep{heintz2020Emerging, frost2021Implementing}. 
These models integrate a variety of healthcare professionals into collaboration, including social workers, registered nurses, non-clinical care managers, physicians, and advanced practice providers. 
By working together, these diverse teams can provide comprehensive and coordinated care, addressing the complex needs of people living with NCDs.
Researchers and practitioners pinpointed continuous monitoring and assessment as one of the eight core elements for comprehensive and collaborative dementia care that enhance outcomes and reduce costs~\citep{super2021Scaling, leeshaggerty2020Recommendations}.
These collaborative care models showed benefits such as fewer behavioral symptoms, enhanced functionality and quality of life, reduced caregiver stress, and lower healthcare expenses~\citep{frost2021Implementing, french2014Healthy, leeshaggerty2020Recommendations}.

In the Chinese context, recent national dementia policies have shifted from early-stage intervention to proactive prevention, with the range of stakeholders having expanded beyond patients to include family members and informal caregivers, as well as broader service entities such as communities, institutions, and social workers, which together support a tiered model of early detection, diagnosis, and intervention~\citep{luo2025evolution}.
Among these stakeholders, family caregivers and healthcare professionals play particularly crucial roles in practice. 
On one hand, family caregiving, especially by adult children, remains the primary form of eldercare in China~\citep{chen2022path, li2022factors}, unlike in countries such as Japan, the UK, or Canada, where community and institutional services provide more extensive social support. 
The awareness and attitudes of family caregivers toward dementia directly influence the likelihood and timing of seeking diagnosis~\citep{wang2019continuum}. 
On the other hand, formal diagnosis of dementia requires multi-faceted assessments, including cognitive screening, neuropsychological test batteries, and biomarker testing, underscoring the indispensable role of specialist clinicians~\citep{van2024computerized, wang2019continuum}.
Despite growing policy attention, systemic challenges remain, including insufficient family and community support, an unprepared healthcare system, inadequate clinical decision support, and others~\citep{lu2025community}.

The rapid development of artificial intelligence (AI) offers significant opportunities to reduce healthcare professionals' workloads and enhance patient outcomes through more accurate diagnosis and personalized treatment plans~\citep{Karalis2024the}. 
In the future, AI may undertake diverse collaborative roles across homes, communities, and healthcare facilities, facilitating continuous care and patient monitoring~\citep{leong2017toward}. 
However, integrating AI into primary care requires careful planning and co-production strategies to ensure successful collaboration~\citep{Astell2021co-production}.
While proposals for AI integration in diagnostics exist~\citep{dolgikh2021collaborative}, challenges remain, particularly in aligning stakeholder needs~\citep{Kuepere2022connecting}.
A systematic review also noted a lack of human-centered design in AI caregiving for NCD patients~\citep{wang2024artificial}. 
Ultimately, a collaborative care model that accounts for NCD patients' unique circumstances is needed, and it will help relieve pressure related to NCDs on healthcare systems.



\subsection{HCI and CAI for NCD Care}
Recent advancements in Human-Computer Interaction (HCI) have shown significant promise in aiding individuals with NCDs in overcoming daily challenges.
New technologies offer novel approaches to support users in various domains, including mobility safety~\citep{wu2016moments}, information security~\citep{mentis2019upside}, physical exercise~\citep{eisapour2018participatory}, emotional communication~\citep{simao2019maty, munoz2021evaluating}, and cognitive reminiscence~\citep{baumann2024mnemosyne}. 
For instance, individuals with NCDs can engage in reminiscence activity through a multimodal approach on the laptop (audio, video, and animation), which made participants highly involved by providing a more relaxed and natural atmosphere than the traditional approach~\citep{gowans2004designing}. 
At the same time, virtual reality provides an immersive, engaging, and effective way to conduct cognitive training~\citep{kim2019effectiveness} and improve physical ability~\citep{neri2017virtual}. 
However, despite their proven benefits, these technologies often present interaction barriers for individuals with NCDs, such as learning how to use joysticks or remembering complex user interface functions, which can hinder adoption and long-term usability~\citep{yusif2016older}.

Conversational artificial intelligence (CAI), powered by natural language processing (NLP) and artificial intelligence (AI), has emerged as a promising solution to support older adults by lowering barriers and simplifying the use of new technologies~\citep{brewer2018accessible}.
CAI mimics human language and behavior to implement certain tasks for the user via a natural language interface, either text-based or voice-based~\citep{abdul2015survey}.
Previous research on conversational robotics in elderly care has highlighted their potential in promoting social interaction while also raising concerns about privacy and over-reliance~\citep{padhan2023artificial, shandilya2024understanding}.
Many studies have explored how CAI supports various aspects of daily life for older adults, including daily care~\citep{padhan2023artificial}, reminiscence~\citep{kang2021momentmeld}, and learning~\citep{entenberg2023ai}.
However, tasks involved in NCD screening differ significantly. 
Unlike social or educational applications, NCD screening focuses on assessing cognitive abilities rather than facilitating interaction or learning.
Besides, users with NCDs may have special needs due to their cognitive conditions.
Previous studies on CAI for individuals with NCDs identified key design requirements, including personalization features like customizing routines, managing calendars, and providing emotional support~\citep{zubatiy2021empowering}.
For instance, AscleAI is a clinical note management system based on large language models (LLMs), designed to help clinicians efficiently record and retrieve patient histories, thereby alleviating some of the diagnostic challenges associated with cognitive decline~\citep{han2024ascleai}. 
In the scenario of NCD screening, CAI has gained research attention for more complex screening tests, particularly those involving conversations. 
For example, a CAI based on ChatGPT-3.5 provided scaffolding for the Grocery Shopping Dialog Task, demonstrating the feasibility of CAI-administered NCD screening~\citep{hu2024designing}. 
While participants in the user study were open to adopting CAI for screening, there remains a lack of understanding regarding stakeholders' acceptance, expectancy, and concerns.



\section{Method}
\begin{figure}[t]
\centering
\revise{
\includegraphics[width=\linewidth]{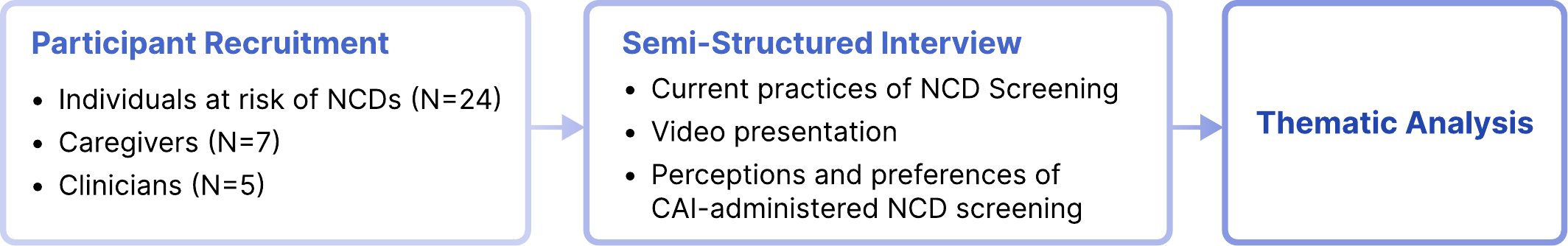}
  \caption{The main procedure of this study.}
  \label{fig:method}
}
\end{figure}

This paper applies a human-centered method by involving key stakeholders to understand their shared and conflicting expectations of CAI for NCD screening in a speculative future. 
Previous research in both primary care~\citep{callahan2006effectiveness, wang2024multi, patel2021engaging, lai2019key} and HCI~\citep{sun2024TechnologyMediated} has pointed out the importance of the collaborative practice of NCD care by involving multi-stakeholders in the process, to produce evidence that matters to patients and families and to impact dissemination and uptake of findings meaningfully.
Building on this, our study employs multi-stakeholder inclusion, involving three participant groups to reflect the key stakeholders that structure the NCD screening pathway.
Individuals at risk of NCDs are directly involved in screening and can provide insights into their firsthand experiences and concerns.
Caregivers, especially informal family caregivers who often assist with health management, play a significant role in interpreting screening outcomes and sustaining technology use over time~\citep{patel2021engaging, luo2025evolution, chen2022path}, making their understanding of everyday care essential to exploring how CAI might be meaningfully adopted.
Given clinicians' role in interpreting screening results and making diagnostic decisions~\citep{mekulu2025reimagining, wang2019continuum}, they provide essential insights into whether and how CAI outputs could support clinical reasoning.
This stakeholder grouping aligns with systematized NCD screening practices in China, where clinicians, caregivers, and individuals at risk of NCDs each play essential roles in initiating, sustaining, completing, and interpreting the screening process~\citep{lu2025community, luo2025evolution, wang2019continuum}.
Moreover, the active collaboration of these stakeholders will be critical when deploying CAI for NCD screening in future practice. 
Issues include understanding the conditions under which CAI would warrant adoption, as well as how much time and cognitive effort stakeholders are willing to invest in the adoption process.
Thus, we conduct semi-structured interviews with the three groups of key stakeholders, as shown in~\cref{fig:method}.
We first investigate current real-world practices of NCD screening during the interviews.
Then, we demonstrate how existing CAI technology facilitates NCD screening by presenting a video to participants that prompts them to speculate about CAI's administration and management of NCD screening in the near future.
Our method is motivated by the design methodology of ``Alternative Presents and Speculative Futures''~\citep{auger2010alternative}.


\subsection{Participants}
In this study, we conducted semi-structured interviews with 36 participants, comprising 24 individuals at risk of NCDs, 7 caregivers, and 5 clinicians.
Participants were recruited according to the following criteria:
\begin{itemize}
    \item \textbf{Individuals at risk of NCDs}: Individuals aged 60 or above~\citep{jia2020prevalence}, with sufficient communication ability to engage in semi-structured interviews. 
    Reflecting the complexity of current screening realities in China, where individuals may be at various stages of seeking or delaying care~\citep{wang2019continuum}, we included individuals with a range of experiences across the screening and diagnosis continuum.
    Participants were included if they had either (a) previously received a formal diagnosis related to cognitive decline, (b) undergone cognitive screening or memory assessment, (c) self-reported noticeable memory decline, or (d) expressed concern about their cognitive status. 
    For those with mild cognitive impairment or more severe cognitive conditions, participation was contingent upon being accompanied by a family caregiver during the session to ensure safety and facilitate communication. 
    \item \textbf{Caregivers}: The primary caregivers for the individuals at risk of NCDs. Inclusion required their active involvement in daily care, medical visits, or the coordination of cognitive screening.
    \item \textbf{Clinicians}: Licensed healthcare professionals who had conducted NCD screening in clinical settings, with a minimum of three years of clinical experience in NCD-related departments (e.g., neurology and geriatrics) within hospital systems. 
\end{itemize}

To ensure a diverse and representative sample, we recruited our participants using two approaches: 
We advertised our recruitment information through social media and community centers to recruit individuals and caregivers by snowball sampling. 
Before the study, participants completed an online recruitment survey.
Based on their responses, we invited individuals and caregivers with diverse backgrounds, e.g., place of residence, experiences of NCD screening, and cognitive health status.
We recruited 24 individuals aged 60 and above (mean = 67.25, SD = 6.39), including 9 males and 15 females.
Six of them participated in the study accompanied by their family caregivers, \revise{while one additional pair of caregiver and individual registered, but the individual did not attend, resulting in us interviewing only that caregiver (CG25)}. 
In total, we interviewed 7 caregivers.
To recruit clinicians, we reached out to 5 specialists from hospitals in Beijing and Tianjin, specifically targeting those with three to over ten years of clinical experience as neurologists.
We focused on professionals from the departments of neurology and neurorehabilitation to ensure both expertise and reliability in their insights.
Detailed demographic information is shown in \cref{tab:participants}, \cref{tab:caregivers}, and \cref{tab:clinicians}. 
For simplicity, in the rest of this paper, we use ``\textit{individual}'' to refer to the individual at risk of NCDs.

\begin{table}[h]
\centering
\caption{Demographic information of the individuals. The individual's cognitive status is based on their previous screening results. NCDs refer to neurocognitive disorders, and SCD refers to subjective cognitive decline.}
\revise{
\begin{tabular}{cccc}
\toprule
Individual ID & Gender & Age & Cognitive Status \\ 
\midrule
P1  & M  & 72 & Mild NCDs  \\ 
P2  & F  & 62 & Unscreened          \\ 
P3  & M  & 60 & Unscreened           \\ 
P4  & F  & 67 & Healthy       \\ 
P5  & F  & 91 & Mild NCDs   \\ 
P6  & F  & 70 & Mild NCDs    \\ 
P7  & M  & 68 & Mild NCDs  \\ 
P8  & F  & 63 & SCD  \\ 
P9 & F  & 62 & Healthy     \\ 
P10 & F  & 63 & Healthy    \\ 
P11 & M & 66  & SCD\\ 
P12 & F & 66 & Unscreened\\ 
P13 & M & 70 & SCD\\ 
P14 & F & 67 & SCD\\ 
P15 & F & 69  & SCD\\ 
P16 & F & 75 & Unscreened \\ 
P17 & F & 70 & Healthy \\ 
P18 & M & 64 & SCD  \\ 
P19 & M & 68 & Healthy  \\ 
P20 & M & 71 & SCD  \\ 
P21 & F & 62  & Unscreened \\ 
P22 & F & 64 & Unscreened\\ 
P23 & M & 62 & Unscreened \\ 
P24 & F & 62 & Unscreened \\ 
\bottomrule
\end{tabular}
}
\label{tab:participants}
\end{table}

\begin{table}[h]
\centering
\caption{Demographic information of the caregivers.}
\revise{
\begin{tabular}{ccccc}
\toprule
Caregiver ID & Gender & Age  & Care Recipient & Relationship \\ 
\midrule
CG1  & F & 67  & P1 & Spouse  \\ 
CG2  & M &  23  & P2 & Mother          \\ 
CG3   & M &  23 & P3 & Father           \\ 
CG5  & F &  28 & P5 & Grandmother  \\ 
CG7  & F & 64 & P7 & Spouse     \\ 
CG21 & F & 54 & P21 & Aunt  \\ 
CG25 & F & 61 & - & Mother  \\ 
\bottomrule
\end{tabular}
}
\label{tab:caregivers}
\end{table}

\begin{table}[h]
\centering
\caption{Background information of the clinicians.}
\begin{tabular}{ccc}
\toprule
Clinician ID & Department & Clinical Experience \\ 
\midrule
C1 & Neurology & 13 years    \\ 
C2 & Neurology & 24 years \\ 
C3 & Neurology & 9 years \\ 
C4 & Neurorehabilitation & 3 years  \\ 
C5 & Neurology & 4 years \\ 
\bottomrule
\end{tabular}
\label{tab:clinicians}
\end{table}

\subsection{Semi-structured Interview Procedure}
The entire interview process was conducted online. 
Before the interview, all participants were invited to review and sign an informed consent form.
The procedure consists of three parts.
All interviews were conducted in Mandarin by coauthors, lasting 40 to 80 minutes through an online meeting platform. 
All interviews were audio-recorded after obtaining each participant's consent and transcribed verbatim in Chinese after the interviews for analysis.

First, the interview included questions on gender, age, city of residence, whether they lived alone, and previous NCD screening results, if any. 
Then, we introduced two screening tests, both known as comprehensive tests covering multiple cognitive domains. 
One was the MoCA~\citep{Nasreddine2005Montreal} because of its popularity~\citep{chen2016validation}.
The other one was the HK-GSDT~\citep{gong2022hongkong} for its plenty of conversational interactions involved in the test and its localization in the Chinese context.
Regardless of their prior NCD screening experience, individuals and caregivers were asked to complete parts of each test. 
This helped them recall or familiarize themselves with the NCD screening tests, facilitating a better understanding of the differences between CAI-administered NCD screening and existing methods and allowing them to express their perceptions more effectively.

The second part focused on the current NCD screening practices. 
The interview questions covered several key topics. 
For individuals and caregivers, we inquired whether they had undergone NCD screening, where it took place, the screening process and methods used, and their understanding of the results. 
For those who had not participated, we explored their reasons and their future intentions regarding NCD screening.
For clinicians, the questions centered around their clinical experience with NCD screening, including the circumstances under which patients visited the hospital, the diagnostic process, the methods they employed for screening, and how they explained the results.

\begin{figure}[h]
  \centering
  \revise{
  \includegraphics[width=\linewidth]{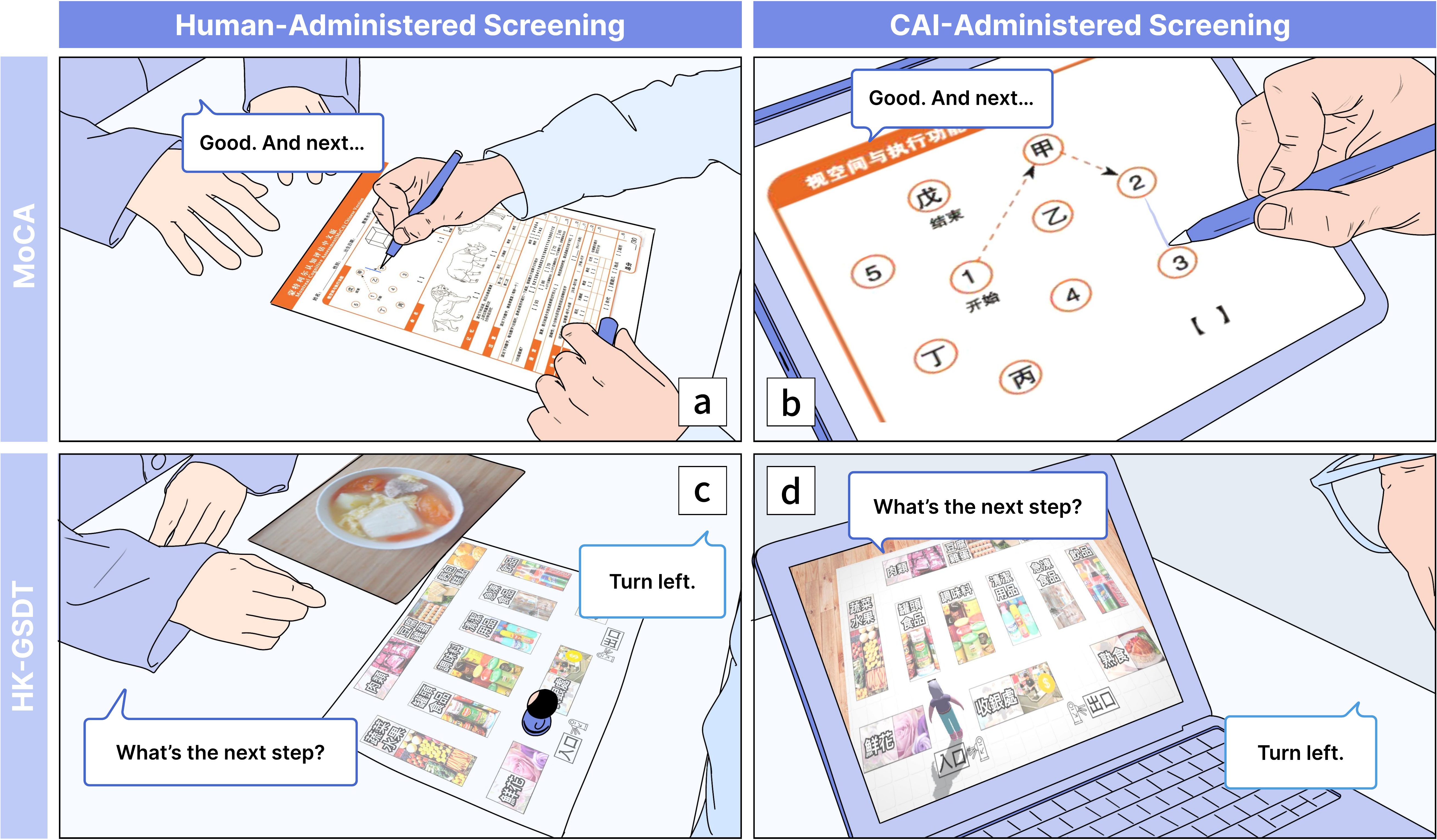}
  \caption{We presented four scenarios for participants. In sketches \textbf{a} and \textbf{c}, a human assessor administers the MoCA test and the HK-GSDT, respectively. In sketches \textbf{b} and \textbf{d}, CAI administers the same two tests.}
  \label{fig:screenshots}
  }
\end{figure}

Then, participants viewed a video demonstrating CAI administering and managing NCD screening, as \cref{fig:screenshots} illustrates.
We composed the video using scenarios of NCD screening administered by a human assessor or a CAI.
This helped the participants relate the concept of CAI-administered NCD screening with familiar technologies to avoid a too-far projection, following the original methodology~\citep{auger2010alternative}.
This visual comparison facilitated a clearer understanding of how CAI-based methods differ from traditional ones, enabling participants to evaluate both approaches more informed.
Based on this video, we guided participants in discussing several key topics. 
We explored stakeholders' perspectives on the acceptance of CAI-administered NCD screening, including expectancies and concerns. 
We also investigated their preferences for adopting CAI for NCD screening, including the location, the device, and the support they might need.
Topics such as privacy and cost were also discussed.
Although most caregivers participated in the interviews alongside their care recipients, they completed their sections independently to ensure comprehensive data collection from distinct stakeholders.
For instance, caregivers were asked to respond to a subset of the same questions that the individuals had answered previously.

\subsection{Ethical Considerations}
The university's Institutional Review Board (IRB) approved this study for research ethics. 
Before we started interviews, we provided participants with detailed information about our study's purpose and background, assured them that their audio data would be used solely for research purposes, and obtained their consent. 
All collected data was anonymized to ensure that no personal identifiers were linked to the data.
Meanwhile, individuals were encouraged to have their caregivers present during the interview, ensuring their comfort and providing necessary support. 
The interviews were positioned as a method to understand participants' experiences rather than a diagnostic procedure, and participants were assured that no medical conclusions would be drawn from their responses. 
Additionally, we used age-friendly communication approaches, avoiding language that could cause unnecessary worry and ensuring a respectful and supportive environment throughout the study.

\subsection{Data Analysis}
We employed a thematic analysis~\citep{braun2012thematic} approach for analyzing the data, with five coauthors participating in the process as coders.
Data from all participants were coded separately, regardless of whether they attended the interview together.
In the open coding phase, the data from each interview session was assigned to two coders to complete the coding and generate initial codes.
Coders frequently met and discussed the codes for alignment and consistency during the iterative process.
Following the open coding stage, we developed an initial code list.
The codes captured participants' NCD screening practices, methods, challenges, and their acceptance, expectancies, and concerns regarding CAI for NCD screening.
We then shifted our focus to broader themes and subthemes, synthesizing related codes into overarching categories.
When no new code or subtheme emerged, the analysis achieved data saturation, suggesting that the data collection and analysis were sufficient~\citep{greg2006many}.
After multiple rounds of discussion, we agreed on a final thematic map comprising three main themes. 

Additionally, we employed user journey mapping, a widely used method in the HCI community~\citep{endmann2016user, bascur2019user}, to understand and design user experience by visualizing interactions and steps throughout the process~\citep{howard2014journey}, which has also been widely applied in healthcare~\citep{ly2021journey, philpot2019creation}. 
Similar to the process proposed by \citet{endmann2016user}, we created a user journey map by analyzing interview data to summarize NCD screening-related activities mentioned by different stakeholder groups, mapping their sequence and locations, and identifying the pain points reported at various stages.
Then, we speculated on how CAI could facilitate routine NCD screening and education. 
By mapping user journeys, we aimed to bridge the gap between the current practices and the speculative future NCD screening with CAI, uncovering how it might address user needs, reduce social stress, and balance the conflicting expectations of stakeholders, ultimately informing a more human-centered design approach.
Thus, the following sections first present the shared and conflicting expectancies of stakeholders regarding the speculated future of adopting CAI for NCD screening.
To better understand the conflicts, we then illustrate the user journey map of their current practices.
Last, we propose design implications for CAI to balance the diverse needs of different users and stakeholders.
\section{Findings}
\subsection{Current Practices for NCD Screening}
Before the speculated CAI, we summarized the current practices of NCD screening based on our interviews.
At first, individuals and their caregivers considered going to the hospital for NCD screening if they experienced obvious symptoms because neurocognitive disorder screening was not yet a common part of individuals' annual health check-ups, according to their responses.
Community clinics offered a convenient option for individuals to seek medical care, such as routine check-ups, basic treatments, and prescriptions. 
According to the clinician, C2, some local community clinics had integrated NCD screening into their annual check-ups, using the Mini-Mental State Examination (MMSE) for brief cognitive tests on older adults.
If abnormal results were detected, the check-up report would advise the individual to seek further examination at a hospital. 
Moreover, individuals and caregivers held a skeptical attitude towards community clinics, believing that they only offered basic services, e.g., \textit{``Community clinics just prescribe medication...They don't have specialized examinations.''}, as P20 mentioned.

Participants preferred not to consider screening until severe symptoms were observed, as P12 mentioned, \textit{``I will not go to the hospital until I have to''}. 
Clinicians (C1-C4) also recalled that most of their patients visited the hospital because they or their caregivers noticed some symptoms. 
Noticeable symptoms, such as memory decline, mental disorders, abnormal behaviors, and reduced capability in daily life, were mentioned.
Aside from experienced or observed symptoms, a family history of NCDs is another motivation according to clinicians and individuals, e.g., P15 and P20 visited the hospitals because of their family history.
Though individuals considered NCD screening, many factors prevented them from taking action.
\textit{``I don't think that we attach much importance to people with dementia. To be honest, if this kind of thing didn't happen in my own family, I wouldn't pay much attention to it either.''} as P14 mentioned, whose father was diagnosed with NCDs at the age of 89.
Individuals and caregivers generally had limited knowledge about when they should screen for NCDs and how.
For example, P6, P11, P18, and P22 had no idea about which institution or department they could screen NCDs at. 
Individuals and caregivers tended to seek out large, reputable hospitals for related tests.
However, visiting these hospitals was inconvenient, and it required effort to make the trip.
\textit{``It's better, more focused... although it's far from us and we have to drive every time... It's quite a hassle.''}, mentioned by P15.
The lack of routine screening within communities, the unawareness or misunderstanding of NCDs, and the inconvenience of visiting the hospital led to a common situation, delayed diagnosis, as acknowledged by C4, that the majority of her patients in the clinic for the first time were already at least with mild NCDs or mild cognitive impairment.

\subsection{Expectancies toward CAI-supported NCD Screening}
\label{sec:shared_expectancies}
Inspired by the Unified Theory of Acceptance and Use of Technology (UTAUT2) \citep{venkatesh2003user, venkatesh2012consumer}, we summarized stakeholders' acceptance of CAI-administered NCD screening in terms of performance expectancy, effort expectancy, social influence, and facilitating conditions, as \cref{tab:acceptance} shows.
First, performance expectancy refers to the functions and services that CAI provides for NCD screening.
Second, effort expectancy is about the estimated effort that the user makes to use the CAI-supported NCD screening.
Third, social influence refers to the extent to which social factors, such as the opinions and expectations of others, affect the user's adoption of CAI for NCD screening. 
Last, facilitating conditions refers to the degree to which individuals believe they can use CAI for NCD screening and access necessary support.

To further analyze how these expectations are distributed across stakeholder groups, we categorized each identified expectation into one of three types: (1) Shared Expectations, i.e., expectations expressed by multiple stakeholder groups, indicating a collective recognition of CAI's value or functionality; (2) Additional Expectations, which refer to expectations emphasized by a specific group beyond the shared ones, reflecting role-specific needs or situated experiences; and (3) Conflicting Expectations, which refer to expectations where stakeholder groups hold opposing views, often due to differences in responsibilities, authority, or ethical considerations.

\begin{table}[t]
\caption{Acceptance of adopting CAI for NCD screening among stakeholders based on UTAUT2~\citep{venkatesh2012consumer}. \ding{52} indicates expectation. \ding{58} indicates additional expectation. \ding{91} indicates conflicting expectations. - indicates unmentioned expectations by the stakeholders.} 
\centering
\begin{tabular}{ m{2cm}  m{5.2cm}   >{\centering\arraybackslash}m{1.3cm}   >{\centering\arraybackslash}m{1.3cm}  >{\centering\arraybackslash}m{1.3cm} }
\toprule
\textbf{Category} & \textbf{Performance Expectancy} & \textbf{Individual} & \textbf{Caregiver} & \textbf{Clinician} \\ 
\midrule
\multirow{9}{*}{\makecell[l]{Performance\\Expectancy}} 
& Comprehensive screening & \ding{52} & \ding{52} & \ding{52} \\ 
\cline{2-5}
& System consistency & \ding{52} & \ding{52} & \ding{58} \\ 
\cline{2-5}
& Emotional support & \ding{52} & -  & \ding{91} \\ 
\cline{2-5}
& User status recognition & - &   \ding{91}  &   \ding{52} \\ 
\cline{2-5}
& Result presentation and management & \ding{52} & \ding{52} & \ding{58} \\ 
\cline{2-5}
& Undesirable result communication & \ding{52} & \ding{52} & \ding{58} \\ 
\cline{2-5}
& Immediate diagnosis & \ding{52} &  \ding{52} & \ding{91} \\ 
\cline{2-5}
& High frequency of screening & \ding{52} & \ding{52} & \ding{91} \\ 
\cline{2-5}
&Facilitating communication & \ding{52} & - & \ding{52} \\ 
\hline
\multirow{2}{*}{\makecell[l]{Effort\\Expectancy}}
& Learning cost & \ding{52} & \ding{52} & \ding{52} \\ 
\cline{2-5}
& Easy setup & \ding{52} & \ding{52} & \ding{58} \\ 
\hline
\multirow{2}{*}{\makecell[l]{Social\\Influence}} 
& Less social stress & \ding{52} & \ding{52} & - \\ 
\cline{2-5}
& Privacy protection & \ding{91} & \ding{52} & - \\ 
\hline
\multirow{4}{*}{\makecell[l]{Facilitating\\Conditions}}
& Availability & \ding{52} & \ding{52} & - \\ 
\cline{2-5}
& Less labor & \ding{52} & \ding{52} & \ding{52} \\ 
\cline{2-5}
& Perceived reliability & \ding{52} & \ding{52} & \ding{91} \\ 
\cline{2-5}
& Confidence in using CAI & \ding{52} & \ding{91} & \ding{52} \\ 
\bottomrule
\end{tabular}
\label{tab:acceptance}
\end{table}

\subsubsection{Shared Expectancies}
First, stakeholders shared expectancies for CAI performance.
P16, P23, CG25, and C2 \textbf{expected CAI-based screening to be more comprehensive} than human-administered screening, specifically, covering multiple cognitive domains through a battery of tests.
This expectation was particularly pronounced among individuals and caregivers, who observed that clinicians in high-volume hospital settings often have limited time to administer extensive assessments—such as MoCA~\citep{Nasreddine2005Montreal}.
Clinicians similarly endorsed this view, highlighting CAI's potential to integrate a more expansive test battery, thereby facilitating a more thorough screening process.
CAI is valuable in \textbf{facilitating communication}, bridging the gap between patients and clinicians, particularly during clinical visits. 
C1 highlighted that the pre-visit information from CAI allowed clinicians to understand the patient's condition better, potentially enhancing the efficiency and focus of the consultation.
Besides, through CAI's ability to present results in an accessible and understandable format, CAI empowers patients to comprehend their cognitive status more effectively, further enhancing their communication with clinicians.
For instance, P14 mentioned, ``\textit{I hope CAI can provide me with valuable information through communication... I also hope it can guide me in communicating with the doctor. For example, when I am in the hospital, what should I say to the doctor about my cognitive issues?''}
This not only reduces the communication burden on clinicians but also enhances patient engagement and understanding, facilitating shared decision-making and proactive cognitive health management.

Under effort expectancy, \textbf{learning cost} refers to the time and effort required for users to learn how to use a new technology or system.
Clinicians, individuals, and caregivers shared a common effort expectancy: they agreed that proper training and education would enhance independent and effective CAI use.
Orientation is the main solution mentioned by participants (e.g., P1, P9, P20, etc.) to enhance their independence in using the new screening tool. 
They emphasized that learning the necessary operational steps and familiarizing themselves with the process could significantly improve their confidence and ability to use the tool independently. 
They suggested such training could be offered by hospitals, professionals, or experienced family members.

Regarding social influence, individuals and caregivers expected CAI could \textbf{reduce social stress} in the current NCD screening in the hospital.
Individuals felt nervous about going to the hospital or seeing a doctor, for example, P5 and P18 mentioned that even the doctor's white coat might cause anxiety.
The resistance was also manifested in the social stigma of NCDs.
Even when clear deterioration was recognized by P18 and his family members, he refused to undergo NCD screening because he did not come to terms with the fact of being ill or aged, especially in front of his kids.
\textit{``It (NCDs) is not fatal, isn't it? Not a big deal...They (the family members) said that I must be ill or something, but I always denied it...to save my face.''}, said P18, who had experienced symptoms but was resistant to NCD screening.
The clinician, C4, also confirmed that this is a typical group of individuals in the clinic.
They would rather not know about their actual situation of cognitive decline to avoid any negative emotions it might cause.
For example, P4 felt rather disappointed about her cognitive decline, even if it was in a non-NCD range.
Two participants, CG5 and P22, felt it was somehow meaningless to screen because no cure has been guaranteed to be effective on NCDs like Alzheimer's disease.
It is considered unnecessary to keep screening if the individuals are already diagnosed with Alzheimer's disease unless there is an effective solution.
CAI has the potential to address these challenges because individuals and caregivers can complete screening at the places where they feel comfortable, instead of the hospital.
Moreover, this may create more opportunities to educate users with NCD-related knowledge to encourage them to face the issue and manage their cognitive health proactively.

Under facilitating conditions, \textbf{availability} is often mentioned by individuals and caregivers (e.g., P4, P23, CG3, etc.) regarding the limited access to clinical services.
Making an appointment with a doctor is rather difficult, but CAI can be available anytime.
P4 mentioned, ``\textit{I have to wait in line to register when I go to the hospital, so I don't want to go. If I could use it at home anytime for testing, that would be much better.}''
Besides, stakeholders (e.g., P6, CG7, C4, etc.) valued the CAI's benefits of \textbf{saving time and labor}.
Completing NCD screening with CAI at home or in the community saves transportation time to the hospital and queuing for the doctor or examinations.
It also improves accessibility for individuals with limited mobility.
From the perspective of caregivers, it also saves their labor if they do not have to take the individual to the hospital, as CG21 mentioned.
It is also agreed by C1-C4.
With the idea proposed by CAI, clinicians could deal with more targeted or urgent cases in the hospital.


\subsubsection{Additional Expectancies}
\label{sec:additional_expectancies}
Regarding \textbf{system consistency}, individuals and caregivers expected a consistent experience, while clinicians highlighted standardized assessments. 
CG7 and P7 recounted inconsistent hospital experiences, such as inadequate consultations with clinicians. 
For instance, P7 received only brief advice from the doctor after waiting extensively—an interaction that frustrated her.
Such inconsistencies amplified their expectations for CAI: they hoped it would maintain consistent empathy and deliver standardized screening interactions, addressing a key limitation in human-provided care~\citep{kerasidou2020artificial}.
Consistency is further recognized as critical to reliable medical assessments~\citep{montemayor2022principle, amann2020explainability}, which shapes clinicians' additional expectations for CAI. 
Unlike human practitioners, who often face challenges with inter-rater reliability and judgment variability~\citep{lucza2015screening}, CAI can mitigate these issues through standardized language and processes, as acknowledged by clinician C3.
Regarding screening \textbf{result presentation and management}, individuals and caregivers (e.g., P4, P6, CG2, CG3, etc.) prioritized concise summaries with plain-language explanations, while clinicians (C1-C5) required granular details—including domain-specific scores, screening logs, and even video recordings of the assessment. 
For example, P4 stated, ``\textit{Just tell me if the score is 60, 70, or 80—avoid using too much medical jargon, as I don't understand it.}''
Clinicians stressed the necessity of detailed data for clinical decision-making: as C2 noted, ``\textit{I need to see the scores across different cognitive domains,}'' and C5 added, ``\textit{I also need the video of the test to assess patient compliance, such as whether anyone nearby is providing prompts.}''
Beyond result presentation, stakeholders highlighted challenges with result management in current care: participants reported losing clinician- or hospital-provided results due to poor record-keeping, while others (e.g., P4) intentionally discarded results due to emotional aversion—she noted, ``\textit{I tore it up and threw it away}'' because she distrusted the results. 
Consequently, all stakeholder groups widely anticipated CAI's ability to automatically record and manage screening results, as this feature would provide a reliable means of accessing critical cognitive data and tracking changes over time.

Another expectancy about performance is \textbf{the CAI's communication skills if the result is undesirable}.
Individuals expressed mixed preferences to be informed directly or not, while clinicians had more considerations. 
As P23 explained, ``\textit{It's okay. I am prepared for any bad news. I heard it's quite common among the aged.}'' 
Conversely, others preferred not to be informed, fearing that such knowledge would cause undue anxiety or harm their well-being.
P2 stated, ``\textit{If the test results are bad, I'd rather not be aware. It would only make me worry.}''
Timing and method of notification were also critical concerns. 
Individuals preferred receiving results immediately after the test, as noted by P2: ``\textit{I want my scores right after the test.}'' 
However, they also emphasized the importance of sensitivity in the delivery, advocating for a communication style that avoids blunt diagnosis and medical jargon, which could be emotionally overwhelming. 
Clinicians reinforced this view, stressing the need for clear but compassionate communication. 
They recommended that CAI focus on delivering actionable suggestions, such as advising users to seek medical attention, rather than providing a definitive diagnosis. 
C3 commented, ``\textit{CAI could provide a prompt indicating that the user may be experiencing early signs of cognitive decline, advising them to seek professional medical consultation as soon as possible.}''
Stakeholders also differed on who should be informed of negative results. 
Individuals preferred that their family members be notified first, as they believed this would facilitate practical support and emotional cushioning. 
For example, P12 remarked, ``\textit{If something is wrong, my family should know first so they can help me deal with it}.'' 
Conversely, caregivers hesitated to share bad news directly with care recipients, citing concerns about causing unnecessary anxiety. 
As CG2 explained, ``\textit{I don't want to bring her too much negative energy.}''
In clinical practice, clinicians typically notify family members or caregivers when results indicate severe cognitive decline. 
C4 explained, ``\textit{When the results are terrible, we usually tell the patient they did okay but avoid sharing the specific scores. We explain the severity to their family members}.'' 

Regarding the expected effort, \textbf{easy setup} refers to the simplicity and quickness with which a user can configure and use a technology or system without complex procedures.
Participants agreed on the home or community setup, while clinicians emphasized more requirements for the non-clinical setup. 
While installing CAI in hospitals for NCD screening may be beneficial to some extent, such as reducing waiting time, the most mentioned usage scenarios by individuals at risk of NCDs and caregivers (e.g., P14, P20, CG1, CG25, etc.) are at home and in community clinics, which are more convenient than hospitals. 
C2 noted that telemedicine is the future of the healthcare system, while C1 pointed out that conducting screenings outside of hospitals could reduce labor costs, improve the reach of early diagnosis, and assist with monitoring interventions and rehabilitation.
However, clinicians emphasized that individuals may encounter interference using CAI for NCD screening at home, which could affect accuracy.
A noisy environment may prevent individuals from correctly understanding the instructions or questions, which could impact their performance. 
Both C3 and C4 stressed the importance of maintaining a quiet environment when using CAI screening at home.
Furthermore, participants' previous experience with technology might significantly influence their device preferences for installing CAI. 
Smartphones and tablets were the most commonly mentioned by participants, such as P3, P7, CG5, etc.
There are two main reasons for this choice.
First, smartphones and tablets are more convenient, especially smartphones, which are portable and easy to carry. 
Second, individuals such as P1 and P13 reported being more familiar with smartphones, whereas other devices (e.g., computers) might pose operational challenges for other individuals.
P10, who used computers regularly at home, preferred to complete the test on a computer.
Additionally, P4, P21, and CG25 suggested that hands-free devices such as voice assistants or smart speakers would be more suitable for older adults, as these devices require no manual operation, support conversational testing, and are convenient and easy to learn.
From the clinician's perspective, C1 proposed two requirements. 
First, the device must accommodate the visual limitations of older users, such as using larger screens, and second, it must address hearing limitations by providing higher volume options.
As C5 suggested, a CAI with a big screen that required no complex operations might increase the individual's compliance.

\subsubsection{Conflicting Expectancies}
\label{sec:conflicting_expectancies}
The conflicting expectancies mainly emerged in performance expectancy and facilitating conditions.
Regarding \textbf{emotional support}, the individuals in our study expected CAI to relieve negative emotions if they encountered difficulties during the screening.
However, clinicians showed concerns that careful and professional design is crucial for having emotional support in CAI so that the reliability of the test can be maintained.
Individuals (e.g., P2, P11, P24, etc.) indicated that emotional support, such as comforting or encouraging, is essential when using CAI for NCD screening.
We outlined two primary reasons for this need: Firstly, emotional support helps reduce the stress and anxiety caused by the hospital environment or performance pressure during NCD screenings, allowing for a more accurate assessment of cognitive abilities.
Individuals and caregivers preferred the CAI approach over traditional human-administered screenings, citing CAI's perceived friendliness and patience, contributing to a more relaxed experience. 
In contrast, the presence of doctors and medical equipment in clinical settings often triggered anxiety and tension, potentially affecting their performance during the screening. 
Secondly, emotional support can enhance individual compliance, particularly when frequent mistakes may lead to frustration, embarrassment, or even early dropout.
Individuals (e.g., P4, P14, P22, etc.) reported they would feel frustration and embarrassment when making frequent mistakes during NCD screening. 
In such cases, emotional support would be essential in helping them manage their emotions, rebuild confidence, and continue with the screening process, preventing premature abandonment. 
As P19 mentioned, no matter how low the score was, if the doctor told her what was positive in the result, she would be happy to accept.
Individuals suggested strategies for emotional support from CAI, including encouragement, affirmation, and the opportunity to try again with friendly avatars.
P20 mentioned, \textit{``If I made a careless mistake, I think it should allow me to try again.''}
Individuals also prefer encouragement and affirmation when making frequent mistakes, such as \textit{``you are performing well compared to your peers,''}and motivational prompts to keep going and improve. 
However, clinicians emphasized that the design must be approached with care and professionalism while such emotional support is necessary. 
Meeting every expectation from individuals and caregivers is not always feasible, as some demands, such as offering additional chances, might compromise the accuracy of the results.
C4 shared an acceptable way from human-administered screenings as reference: experienced assessors may use standardized phrases such as \textit{``It's okay. This question is a bit hard''}, and \textit{``Let's move on to the next one.''}

Aside from the conflicting expectations for emotional support, caregivers had disagreements with clinicians' expectations for \textbf{user status recognition}.
The individual characteristics and usability challenges among older adults can vary a lot, which made the clinicians expect CAI to recognize user status, such as checking the individual's hearing and reading capability to adjust strategies for screening.
Clinician C4 emphasized the importance of CAI recognizing individuals' hearing, vision, language preferences, and other status information to ensure accurate assessment of cognitive abilities. 
For instance, CAI would need to account for sensory impairments, such as hearing or vision loss, prevalent among older adults, as noted by C1 and two participants. 
Failure to do so could significantly reduce its usability and the validity of screening results.
Caregivers agreed with the need to improve usability, but CG2 and CG5 raised privacy concerns about the caregiver's personal data involved in the process.
Then, participants had more diverse expectations for user status recognition, such as the user's status of language comprehension. 
Participants (e.g., CG2, CG5, C1, C4, etc.) expressed concerns that CAI might struggle to understand dialects, a limitation that could undermine its effectiveness in certain populations. 
Furthermore, individuals with limited literacy or familiarity with technology may struggle to interact effectively with CAI or the digital devices hosting it. 
This highlights a broader accessibility issue, which must be addressed to ensure equitable use of the technology.
Lastly, poor compliance among some individuals poses an additional barrier. 
Clinician C4 shared that some patients, particularly those who believed they were healthy, might resist cooperating during the screening process. 
As C4 explained, \textit{``They feel like it is meaningless to ask them those questions. When you ask these questions, they think you are humiliating them.''} 
Similarly, individuals with major NCDs often refused to cooperate, possibly due to their inability to complete the required screening tasks.
These behavioral and cognitive challenges further emphasize the need for adaptive and empathetic design in CAI systems.
Individuals and caregivers (e.g., P14, CG25) would be eager to get an \textbf{immediate diagnosis} and medical advice from CAI screening. 
However, clinicians believed this should be strictly prohibited because the clinician must deliver the diagnosis based on the results of multiple formal examinations. 
Clinicians recommended that CAI refrain from diagnosing but provide general guidance to seek professional medical evaluation. 
As C3 noted, ``\textit{Simply using some cognitive test scales to inform patients of their cognitive results is insufficient. Additional diagnostic tools like PET scans or fMRI are often employed in clinical settings for a more accurate diagnosis}.''
This view is consistent with previous research, which emphasizes that AI systems should support, but not replace, clinical decision-making, particularly regarding delivering diagnosis~\citep{amann2022explain}.
Therefore, CAI should only provide suggestions for individuals and caregivers to seek medical advice.

Additionally, \textbf{frequency of NCD screening} with CAI revealed a conflict between clinicians' recommendations for routine intervals and individuals' diverse preferences, driven by varying levels of anxiety, understanding of NCDs, and perceived necessity for screening.
In practice, the frequency of screening is adjustable based on individuals' status. 
For individuals at risk of NCD, clinicians generally recommend that they keep routine screening at an interval of 6-12 months, even though the previous results have indicated that they do not have NCDs. 
Participants (e.g., P17, P22, CG21, etc.) hoped to be aware of individual cognitive states, acknowledging that cognitive abilities would change continuously with age.
They understood that screening was not a one-time task, and staying informed about their condition allowed for early intervention to maintain a good quality of life and reduced the burden on their caregivers. 
However, some participants (e.g., P1, P19, CG2, etc.) expressed anxiety about NCDs and wanted to increase the frequency of screening, such as every three months, monthly, weekly, or even daily.
Nevertheless, as clinicians mentioned, it might lead to inaccurate results due to learning effects. 
P19 misunderstood the screening, hoping that frequent practice on the same test could improve their scores on future tests. 
Their intense concern about their cognitive state led to expectations that conflicted with the clinicians' recommended frequency that can ensure the tests' validity.
CG3 and P7 held a contrasting view: if the results were normal, they would temporarily refrain from routine NCD screenings. 
They believed they could roughly assess their own or their care recipient's condition and would only do screening again if symptoms were noticed.
Clinicians suggested the screening frequency should be adjusted according to the stage and treatment goals for individuals diagnosed with NCDs and receiving medical interventions. 
C4 advised that patients undergoing intervention complete follow-up screenings every 1–3 months to evaluate treatment effectiveness.
In early-stage patients, monthly screening should be appropriate when medication or treatment plans require adjustment based on outcomes. 
C2 emphasized that the management of NCDs should be continuous, with regular screening necessary to monitor both pharmacological and non-pharmacological interventions. 
However, patients with NCDs often failed to adhere to follow-up schedules, leading to low rates of revisits, as noted by C5.
On the other hand, it is unnecessary to keep routine NCD screening for those patients with major NCDs because they already receive the maximum dose of medical intervention or cannot complete any NCD screening tests. 
The caregiver, CG5, whose care recipient had moderate NCDs, mentioned that since there was no effective treatment, routine screening for diagnosed patients seemed less meaningful, leading to low motivation; however, acute cognitive decline, potentially caused by other diseases, was still important and could be detected through routine screening.
As C4 mentioned, regular screening could help detect abnormal disease progression and assist doctors in identifying the underlying causes.

Regarding social influence, \textbf{privacy protection} is important, especially when CAI is deployed in private spaces such as the home.
Because CAI collects personal data, such as the recordings of conversations between CAI and the user, as well as their NCD screening data reflecting their cognitive status.
Individuals and caregivers hold conflicting views on privacy protection.
It was surprising that participants had little privacy concern about using CAI for NCD screening.
P6 mentioned, ``\textit{This is just about doing tests and answering questions. It doesn't involve any privacy problems, so I don't worry about it at all.}''
CG3 thought that the personal information CAI collected should be insensitive, ``\textit{It's just some quiz.}'' as he mentioned.
However, the caregivers are sometimes of younger generations, and they have better privacy awareness, e.g., CG5, who took care of her grandmother, thought that CAI should not collect information from the caregiver or other persons around.

In facilitating conditions, \textbf{perceived reliability} refers to users' confidence in CAI's ability to provide accurate, consistent, and trustworthy screening results. 
Individuals and caregivers trusted the CAI's screening results to different extents.
However, clinicians considered it to be only a suggestion on whether medical consultation was needed.
These individuals were familiar with various AI products in daily life and generally trusted technical tools like CAI. 
For instance, P16 stated, \textit{``Digital things do not lie to anybody.''}
However, this trust was contingent on evidence supporting CAI's effectiveness and accuracy, as emphasized by individuals.
Some participants thought that CAI was perceived as less reliable compared to human assessors or doctors.
P9 and P10 thought that they were supposed to complete screening with CAI in the presence of a human doctor or professional.
Compared to CAI, participants thought the human doctor could better observe the patient and answer any kind of questions during the screening.
Or, as C4 mentioned, the human doctor could be a supervisor in the present, and some patients would be more concentrated.
However, clinicians adopt a more cautious stance, acknowledging CAI as a useful tool for early screening and for prompting individuals to seek further medical consultation if needed. 
They stressed that a formal NCD diagnosis should combine multiple tests and clinical judgment.
Last, individuals and caregivers held conflicting views on \textbf{confidence in using CAI}.
Individuals (e.g., P4, P6, P23, etc.) were confident in using CAI for NCD screening, though some also expected step-by-step natural language guidance during the process to help them complete the test. 
However, some caregivers of individuals with major NCDs lacked confidence in their care recipients' ability to use CAI. 
For instance, the caregiver CG2 noted, ``\textit{If they face a problem, they wouldn't know how to solve it. It would be helpful if there were someone they could contact, like an online assistant, to guide them step by step.}''
This difference might stem from variations in literacy and cognitive abilities.
The clinician, C4, thought that most of the patients should be able to complete screening with CAI, except for a small part that might need reevaluation.
Therefore, clinicians recommended that CAI be adopted for individuals who did not have major NCDs yet as an early screening method.

\section{Discussion}
\subsection{User Journey of NCD Screening}
\begin{figure}[h]
  \centering
  \revise{
  \includegraphics[width=\linewidth]{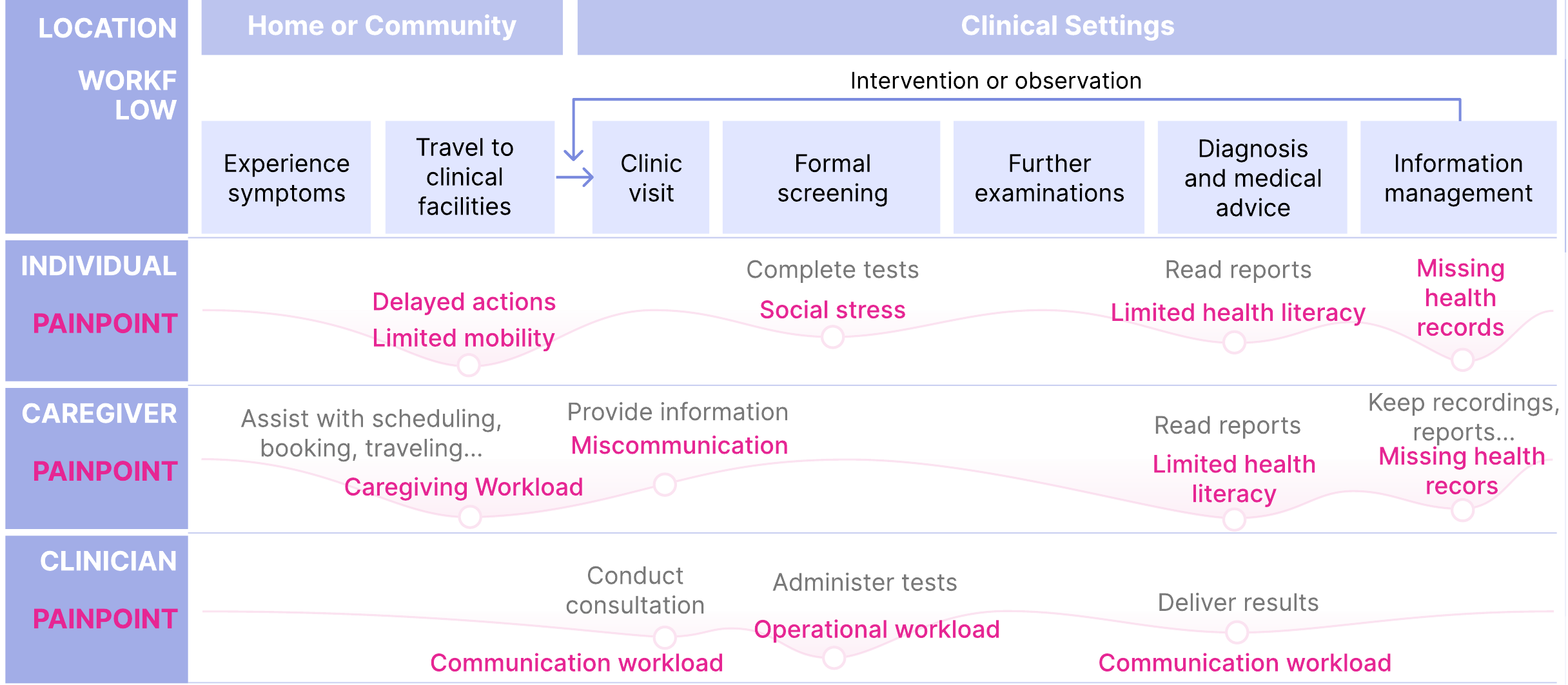}
  \caption{The original user journey map of the hospital NCD screening and its pain points. Individuals and caregivers go to the hospital after experiencing symptoms, usually leading to a delayed diagnosis. Other issues, such as individuals' social stress, the workload of caregivers and clinicians, and inefficient communication, might also hinder accessible NCD screening.}
  \label{fig:original_flow}
  }
\end{figure}

To better understand the conflicts among stakeholders, we discussed the user journeys of the current practices for manual NCD screening in the hospital, as~\cref{fig:original_flow} shows based on our interview.
From stakeholder interview accounts, we found that currently delayed NCD screening or diagnosis could be attributed to multiple factors, such as skepticism about the reliability of having community physicians deal with NCD issues.
The literature echoed this with evidence that the clinician's lack of geriatric experience and individuals' doubt about the health care system are substantial obstacles to timely NCD recognition~\citep{chodosh2004physician, Henley2023happy}.
Moreover, it is a typical attitude for individuals to consider NCDs as natural phenomena instead of a disease that needs treatment~\citep{bradford2009missed, Henley2023happy}, especially Asians~\citep{JONES2006asian}.
\citet{JONES2006asian} also mentioned that Asian individuals' information sources about NCDs are scattered from personal experience and the media, leading to their confusion or misinterpretation about NCDs.
Other practical factors, such as clinicians' office time burden and social stigma about NCDs among stakeholders, might also lead to late or missed NCD detection~\citep{bradford2009missed}.

The caregiver often assists the individual before the hospital screening, e.g., by making an appointment or assisting transportation, which also causes inconvenience, aligning with the diagnostic barriers identified in the literature~\citep{ortiz2000barriers, charles2025navigating}.
When consulting the clinician, caregivers provided information such as family history or symptom description.
However, \citet{vick2018speak} identified communication challenges in the primary care visits involving individuals with NCDs and their family companions (usually as caregivers).
For example, role confusion about who should speak, decide, or be the ``real'' patient.
Sometimes, clinicians cannot tell whether to trust the patient's answers or the companion's corrections.
According to our participants, the clinician or another professional assessor administers the NCD screening tests in the hospital, which might bring a significant workload.
For example, the commonly used MMSE in clinical practices usually takes 8-15 min for each subject~\citep{Mitchell2017mmse} and physicians in China dealt with 40 patients per day on average~\citep{Zhang2018salary}.
Moreover, our findings show that individuals frequently feel potential distress about the possibility of dementia and social stigma, aligning with the literature~\citep{bradford2009missed, JONES2006asian}.
After the screening and examinations, the clinician delivers the result with brief explanations to the individual and the caregiver.
According to our participants, it is common for the individual and the caregiver to have insufficient comprehension of the result due to the time limit caused by the high patient volume in the hospital.
This is echoed by the health professionals' expectations for an automated tool for NCD risk communication~\citep{Naismith2022characterising}.
Back home from the hospital, the screening recordings or reports are not well-kept due to the lack of management.
Though digital health records had been promoted, test result management faced challenges such as information overload and coordination difficulty across distributed care~\citep{micheal2019test}.
In summary, our findings echoed multiple challenges in current NCD screening practices as mentioned in the literature, including the limited awareness of NCDs, screening inconvenience and workload, social distress, and lack of information coordination.

To envision how CAI may transform the current practice of NCD screening, we constructed a speculative user journey mapping (\cref{fig:collaborative_framework}) based on interview findings and literature insights.
\revise{In response to the pain points identified in the current practice (\cref{fig:original_flow}), the conceptual CAI-supported pathway is envisioned to enable a more proactive, coordinated screening process.
Distinct from today's one-off, hospital-centred screening model, the proposed CAI-enabled framework conceptualises a recurring routine screening loop.
As anticipated by stakeholders, ubiquitous 24/7 support and standardised assessment delivery within home and community settings could alleviate perceived social stress related to hospital attendance among older individuals~\citep{patterson2022world}, and may also ease the anticipated caregiving burden on informal caregivers.
In hypothetical scenarios where abnormal indicators are identified, CAI is conceptualised to deliver timely guidance, supporting individual users and caregivers in result interpretation and prompting appropriate subsequent medical consultation.
Such regular, CAI-facilitated screening is theorised to improve the timeliness of cognitive assessment and support more effective patient–clinician communication.
Clinicians are also expected to gain benefits from this conceptual workflow: continuously accumulated screening records may provide insightful references regarding patients' long-term cognitive trajectories, especially post-rehabilitation, which could inform more integrated care coordination, intervention outcome evaluation, and personalised care plan adjustments~\citep{callahan2006effectiveness, frost2021Implementing}.
Despite the promising potential and stakeholder-endorsed advantages of CAI-supported NCD screening outlined in this speculative process, novel challenges remain inherent across stages, including system learning burdens, privacy vulnerabilities, and concerns over reliability.
These tensions, derived from stakeholder accounts and synthesised within our conceptual mapping, subsequently underpin the design implications presented in the following sections.
}


\subsection{Design Implications for CAI in NCD Screening}
\begin{figure}[h]
  \centering
  \revise{
  \includegraphics[width=\linewidth]{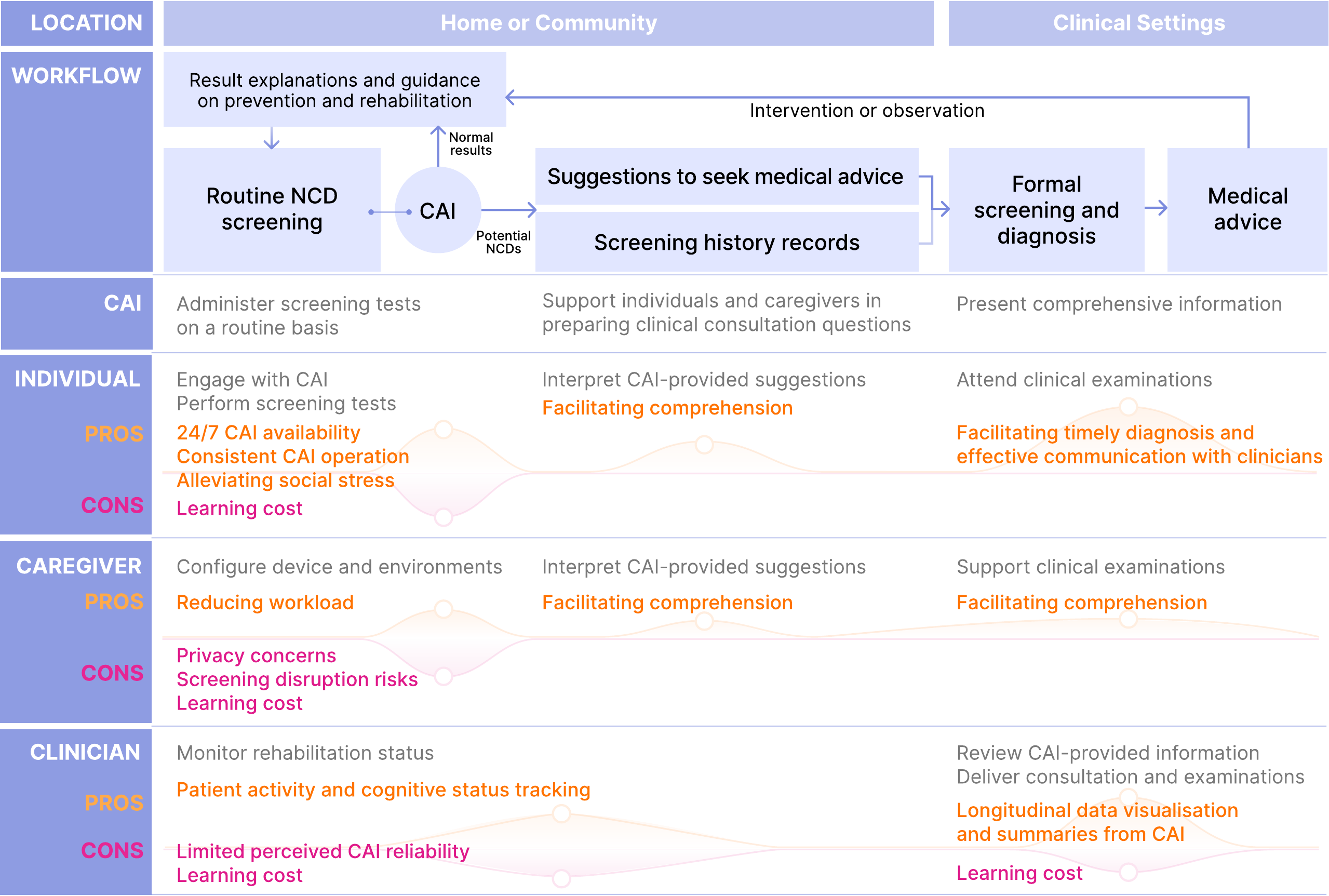}
  \caption{
  The speculative user journey map illustrates routine CAI-supported NCD screening conducted within home and community settings.
  Within this conceptual workflow, CAI is envisioned to raise public awareness of NCDs and deliver timely prompts to facilitate appropriate medical consultation when needed.
  In clinical settings, the proposed CAI system is conceptualised to supply clinicians with holistic patient information, including accumulated prior screening records.
  }
  \label{fig:collaborative_framework}
  }
\end{figure}

Previous Human-Computer Interaction (HCI) studies on CAI for NCD screening mainly focused on conversational skills for test administration~\citep{ding2022talktive, hu2024designing}.
In this work, we look into the full process of NCD screening by illustrating the user journey mapping of the current practices (\cref{fig:original_flow}) and comparing it with the mapping of CAI-facilitated NCD screening (\cref{fig:collaborative_framework}).
Meanwhile, NCD screening, as well as NCD care, is a collaborative process~\citep{callahan2006effectiveness, heintz2020Emerging, frost2021Implementing} and we advocate designing CAI as a coordinator among stakeholders.
So, we highlight the human-centered mindset in designing CAI for NCD screening from a multi-stakeholder perspective.

\subsubsection{Improve Transparency and Build Trust}
Based on interview findings that stakeholders prioritize CAI's institutional legitimacy for building trust, the CAI for NCD screening should disclose its source, such as verification by a professional research institute or a major public hospital.
For instance, a previous study by \citet{low2021attitudes} suggested that older adults tend to seek health information from medical professionals or institutes with authority.
It is advisable to design a CAI that may disclose information related to its developer or the evidence to support its validity.
Stakeholders' perceptions of CAI's reliability vary across interviews.
According to the interview, clinicians are more cautious about adopting CAI due to concerns that individuals may interpret CAI's information as a formal diagnosis.
Beyond taking care to avoid this misinterpretation when interacting with individuals and their caregivers, it is equally important to inform clinicians that CAI will appropriately communicate screening results to users. 
As suggested by \citet{asan2020artificial}, this clarity improves its behavioral transparency and helps clinicians build trust in CAI.
Clinicians have the most authority in the collaborative process of NCD screening, so their attitudes are important to promote the adoption of CAI for NCD screening.

\subsubsection{Adopt and Contextualize the CAI at Home or in the Community}
According to our participants, stakeholders commonly expect that an always-available CAI at home or in the community may provide convenience for routine NCD screening.
Especially for individuals who have limited mobility or are resistant to going to the hospital for screening, it may reduce their concerns and social stress with the out-of-hospital screening by the CAI.
The COVID-19 pandemic accelerated the trend of digitizing NCD management out of the hospital~\citep{rohr2021impact}.
Prior work indicated that older adults living in the community already adopted CAI for a range of activities~\citep{anneliese2024what} and exhibited distinct perceptions and preferences toward CAI~\citep{du2025ai}.
Moreover, our participants mentioned the social stigma of NCDs, a typical issue in NCD screening and diagnosis~\citep{low2019communicating, patterson2022world}.
Involving individuals at risk of NCDs in routine screening practices and providing them with related knowledge about NCDs may facilitate their understanding of NCDs and reduce social stigma~\citep{patterson2022world}.
Therefore, designing CAI for the context of home or community should be prioritized to achieve routine NCD screening and facilitate NCD education.
On the other hand, the sound environment and the conversations might be quite complex in a context such as home~\citep{porcheron2018voice}.
Especially if CAI relies on voice interaction, contextual information such as the speaker's identity should also be carefully considered.
Therefore, to meet the user needs of adopting CAI at home or in the community, designers should consider the physical environment, such as the device and its size, while addressing the clinicians' concerns about the factors that might confound the screening result, such as interruption by irrelevant personnel.

\subsubsection{Provide Orientation to the Target Users}
The individual and the caregiver need appropriate orientation on the use of CAI to ensure the screening effectiveness, according to the interview.
For example, caregivers may provide necessary assistance during the screening, but they also need to be aware that providing excessive support to enhance the individual's test performance may be considered an intervention, leading to invalid screening results. 
Users need to follow proper procedures to prevent operational errors from affecting the reliability of the results.
To effectively integrate CAI into daily lives, recurrent training and ongoing support are essential, similar to other social technologies for NCD patients and their caregivers~\citep{unbehaun2020social}. 
Notably, the primary target users should be healthy individuals at risk of NCD or those with mild NCDs.
\revise{CAI could be particularly beneficial for individuals with mild NCDs undergoing interventions, as it may facilitate frequent monitoring and timely medication adjustments.}
For patients with more severe conditions, CAI may struggle to accurately measure cognitive level for their limited engagement~\citep{saxton2005development}, or the screening tests may become unapplicable~\citep{choe2008severe}.
Therefore, designers should focus on designing CAI for the target groups, healthy or mild-NCD individuals, as well as the clinicians, because they might also use CAI to access their patients' information sometimes.

\subsubsection{Support within a Legitimate Scope}
Stakeholders have conflicting views on how much support CAI should provide during the screening.
According to the interviews, individuals and caregivers prefer CAI for its consistent display of empathy as a part of emotional support. 
Beyond standardized guiding language, individuals need support tailored to their emotional state to reduce stress and anxiety while also enhancing compliance. 
To achieve this, a CAI may be equipped with speech emotion recognition and empathetic feedback~\citep{hu2022acoustically}, or a friendly avatar~\citep{wonggom2019effectiveness}.
Meanwhile, clinicians favor CAI due to its consistency in NCD screening, particularly its standardized language.
This is aligned with a previous study in which a CAI utilized ``\textit{just enough}'' scaffolding skills to administer NCD screening tests~\citep{hu2024designing}.
Additionally, \citet{hu2024designing}'s method of following the professional experts' practices is also applicable.
Thus, designers may address the conflict by limiting the emotional support provided by CAI within a legitimate scope, as demonstrated by human professionals, instead of as much as possible.

\subsubsection{Personalize Information for Different Stakeholders}
Our interview indicates that communication conflicts exist between individuals, caregivers, and clinicians.
Individuals and caregivers currently struggle to understand and manage NCD screening results, even when clinicians provide explanations and printed reports. 
Since patients tend to seek personalized information related to their own situation rather than mere biomedical facts~\citep{ong1995doctor}, the future design of CAI can integrate personalized explanations of diagnostic results and tracking individual changes to enhance patients' understanding.
While individuals and caregivers expect an overall screening result and recommendations from CAI, clinicians need more detailed information about the screening process to validate the results. 
To address this issue, providing clinicians with audio or video recordings of CAI's screening process could offer valuable insights beyond the results, helping to identify potential abnormalities and supporting the final diagnosis.

Clinicians and previous studies highlighted that the way a diagnosis of NCDs is communicated appropriately depends on several factors, including the patient's condition, the healthcare and social care systems (such as insurance coverage and intervention accessibility), and cultural norms surrounding NCDs, such as stigma~\citep{low2019communicating}. 
It remains a challenge for CAI to deliver results with the same compassion and flexibility as a clinician. 
For example, since NCDs are usually irreversible~\citep{zverova2019Clinical}, the individual may have negative emotions, noticing that the cognitive ability keeps declining.
Besides being empathetic, CAI may provide advice on prevention or cognitive rehabilitation for the individual to improve positive thinking.
If suspicious cognitive decline is detected, CAI should instantly recommend seeking specialist clinicians rather than providing a direct diagnosis, allowing specialists to conduct more thorough evaluations and deliver results in a manner tailored to the patient's condition.
In summary, CAI should tailor the screening result delivery to meet the needs of different users, considering their familiarity with NCDs.
For instance, clinicians and people with greater knowledge of NCDs should receive detailed information, while individuals and caregivers who are less familiar with NCDs should be provided with simplified results and clear explanations.
Empathic language styles should be considered.
More importantly, CAI only provides suggestions to seek medical advice, instead of any form of diagnosis.

\subsubsection{Design Alternative Screening Tests}
Due to a lack of knowledge about NCDs and screening tests among caregivers and individuals, or driven by anxiety, individuals tend to increase the frequency of screening, as mentioned in \cref{sec:conflicting_expectancies}.
However, the clinicians suggested that CAI use different tests from a battery of standard tests to avoid learning effects and invalid results.
The MoCA~\citep{Nasreddine2005Montreal} and the MMSE~\citep{flstein1983MMSE} were the most frequently mentioned in interviews.
Complementary to these, other recently validated tests that incorporate real-life scenario simulations familiar to participants, such as the Hong Kong grocery shopping dialog task~\citep{gong2022hongkong}, could also be integrated into the CAI's test battery.
Additionally, CAI may add intervention or training features such as music therapy, reminiscence, cognitive games, and relaxation, mentioned by \citet{cruz2020social}.
Thus, to meet stakeholders' different needs, designers may consider adding different clinically validated tests or providing individuals with additional modules for cognitive training or intervention.

\subsubsection{Protect Caregiver Privacy}
Interview findings revealed divergent privacy concerns toward CAI across generations: younger caregivers expressed greater privacy worries, while older individuals showed less concern. 
This aligns with prior research indicating that cultural values and social norms shape privacy awareness~\citep{cao2008user}, and that older Chinese adults generally have lower digital privacy awareness and limited technical knowledge to protect it~\citep{tao2016elderly}.
When using CAI, it is crucial to emphasize potential privacy risks to both caregivers and individuals, ensuring that they are informed of these risks before collecting any private data. 
Obtaining consent from both parties can benefit them both: it not only raises privacy awareness among individuals and fosters trust in CAI's privacy protections but also allows younger caregivers, who may have greater concerns, to assume a protective role in safeguarding the individual's privacy.
Furthermore, privacy-preserving technology could be applied in data collection of CAI, such as the 3D wireframe reconstruction~\citep{Kunchala_2023_WACV}.

\subsection{Generality of Findings}

\subsubsection{Structural Challenges Shared Across Healthcare Contexts}
This study was conducted in urban settings in China, with participants primarily recruited from city-based communities and public hospitals, which may limit the direct transferability of the findings to rural or severely under-resourced areas. 
Nevertheless, several challenges identified in this study, such as delayed NCD screening, fragmented care pathways, and unclear responsibility for initiating screening, reflect structural characteristics of healthcare delivery rather than culture-specific practices. 
Prior research has characterized NCD screening and care as collaborative yet often poorly coordinated processes, in which preventive activities are weakly integrated into routine care pathways~\citep{callahan2006effectiveness, heintz2020Emerging, frost2021Implementing}. 
This suggests that these challenges may also be relevant to healthcare systems with comparable institutional arrangements.

\subsubsection{Generalization of CAI as a Coordinator in NCD Screening}
A key design implication that generalizes beyond the Chinese context is the role of CAI as a coordinator in the NCD screening process. 
Participants consistently envisioned CAI as supporting coordination among individuals, caregivers, and clinicians by structuring screening workflows, mediating information exchange, and facilitating timely escalation to clinical care. 
This coordination need arises from the inherently distributed and multi-stage nature of NCD screening, which has been widely documented in chronic care and early detection contexts. 
Prior research has shown that breakdowns in coordination across home, community, and clinical settings are a major barrier to effective chronic disease management, and that digital systems designed to support coordination can improve continuity and accountability across stakeholders~\citep{callahan2006effectiveness, frost2021Implementing, heintz2020Emerging}. 
These findings suggest that designing CAI to act as an explicit coordinator, rather than a standalone screening tool, is broadly applicable to healthcare contexts where early detection depends on multiple actors and settings.

\subsubsection{Generalization of Transparency- and Trust-Oriented CAI Design}
Another set of design implications that generalizes across contexts concerns transparency and trust-building in CAI-supported screening. 
Our findings indicate that stakeholders' willingness to adopt CAI depends on clear communication of system scope, data provenance, and limitations, particularly to avoid screening outputs being misconstrued as formal diagnoses.
This concern is consistent with prior HCI and medical informatics research showing that lack of transparency and unclear responsibility boundaries undermine trust in AI-assisted healthcare systems~\citep{asan2020artificial}.
Studies focusing on older adults further suggest that perceived institutional legitimacy plays a critical role in shaping trust in digital health technologies~\citep{low2021attitudes}. 
While the specific sources of authority that confer legitimacy to CAI may vary across healthcare systems, the underlying design principle—making system intentions, evidentiary basis, and responsibility boundaries explicit—is transferable to other contexts where AI is introduced into health-related decision-making.

\subsubsection{Context-Dependent Generalization of Interaction and Support Strategies}
In contrast, design implications related to interaction style, emotional support, and result communication are less directly generalizable and require contextual adaptation. 
Prior research has demonstrated that expectations around diagnosis communication, stigma, and emotional support in NCD contexts vary across healthcare systems and sociocultural settings~\citep{low2019communicating, patterson2022world}. 
At the same time, the broader design challenge of balancing emotional support with clinical appropriateness and clearly bounded system responsibility has been identified across CAI-assisted screening and assessment systems~\citep{hu2024designing, asan2020artificial}. 
This suggests that while specific interaction strategies should be locally adapted, the need to design CAI that is sensitive to stakeholder roles and responsibility boundaries represents a generalizable concern across NCD screening systems.

\section{Limitations and Future Work}
The first limitation of this study is the participant sampling.
Though the analysis achieved data saturation~\citep{greg2006many}, the sample size might affect its validity in a larger-scale context. 
To improve the quality and credibility of qualitative analysis, \citet{patton1999enhancing} proposed triangulation, suggesting future work to explore different data sources (e.g., conducting interviews in different cultural contexts) and different methods (e.g., creating questionnaires to survey on a larger scale to collect quantitative data).
Second, our study highlights the human-centered method in design CAI-administered NCD screening, but the interview only covers the early stage of design.
Future work may consider further exploration that involves multiple stakeholders as evaluators of the prototype design, such as the previous works by \citet{ding2022talktive, hu2024designing}.
More importantly, the long-term effectiveness of CAI in the user's house or in the community should be assessed, e.g., by evaluating the retention of CAI for routine NCD screening.
While this study primarily focused on conversational AI, future work could explore the use of speculative design approaches, as suggested by ~\citet{dunne2024speculative}, to envision and investigate other interactive forms and future interfaces for NCD screening.
This may include interfaces in formats and devices beyond traditional screen displays, making the system more engaging, socially acceptable, and aligned with preferred futures.


\section{Conclusion}
In this study, we invited key stakeholders related to neurocognitive disorder (NCD) screening, clinicians, caregivers, and individuals at risk of NCDs, to share their envisioned near future of NCD screening with conversational AI (CAI) through semi-structured interviews.
We found that stakeholders expressed shared expectancies, such as completing routine NCD screening with CAI at home or in the community, instead of the manually administered and hospital-based screening in the current practice.
This was perceived by stakeholders to reduce or avoid the social stress of individuals, which usually prevents them from seeking medical consultation and leads to delayed diagnosis.
We further identified divergent views among stakeholders.
For instance, individuals and caregivers looked to CAI for support on emotional issues associated with NCDs, while clinicians expected CAI for consistent and standardized screening administration.
Then, we proposed actionable implications of designing CAI to resolve these conflicts and address the diverse expectations of stakeholders, and to facilitate routine screening and proactive management of NCDs.
Finally, we discussed the generalizability of our interview findings, including structural healthcare challenges (e.g., delayed NCD screening, fragmented care pathways) that were identified as shared across comparable healthcare systems, as well as generalizable CAI design implications related to its role as a screening coordinator and transparency- and trust-oriented design.

\section*{Acknowledgements}
The work described in this paper was partially supported by three grants from the Research Grants Council of the Hong Kong Special Administrative Region, China (Project Reference Numbers: AoE/E-601/24-N, T45-407/19N, and 16207923).
as well as by the Global STEM Professorship Scheme of Hong Kong. 
The authors also wish to express sincere gratitude to Prof. Mo Li and Minghui Liu for their special support.
We sincerely thank all participants for their time. 
Special gratitude goes to clinical professionals for generously sparing time amid their busy schedules, individuals at risk of NCDs for their commitment, and their family members for invaluable support, without which this research would not have been possible.
We gratefully acknowledge the large language models, including Doubao‑Seed‑2.0 and GPT-4o, for providing linguistic refinement and polishing of this manuscript, without altering the original meaning, arguments, or research content.







\bibliographystyle{elsarticle-harv} 
\bibliography{ref}



\end{document}